\documentclass[%
 aip,
 amsmath,amssymb,
 reprint,%
]{revtex4-1}

\usepackage{graphicx}
\usepackage{dcolumn}
\usepackage{bm}

\usepackage[utf8]{inputenc}
\usepackage[T1]{fontenc}
\usepackage{mathptmx}
\usepackage{etoolbox}
\usepackage{txfonts}
\usepackage{adjustbox}
\usepackage{color}
\usepackage{rotating}
\usepackage{tabularx}
\usepackage{cases}
\usepackage[colorlinks=true, allcolors=blue]{hyperref}
\usepackage[colorinlistoftodos,prependcaption]{todonotes}

\makeatletter
\def\@email#1#2{%
 \endgroup
 \patchcmd{\titleblock@produce}
  {\frontmatter@RRAPformat}
  {\frontmatter@RRAPformat{\produce@RRAP{*#1\href{mailto:#2}{#2}}}\frontmatter@RRAPformat}
  {}{}
}%

\makeatother
\begin{document}

\preprint{AIP/123-QED}

\title{Extreme multistability in symmetrically coupled clocks}

\author{Zhen Su}
\affiliation{Potsdam Institute for Climate Impact Research, 14473 Potsdam, Germany}
\affiliation{Department of Computer Science, Humboldt-Universität zu Berlin, 10099 Berlin, Germany}

\author{Jürgen Kurths}
\affiliation{Potsdam Institute for Climate Impact Research, 14473 Potsdam, Germany}
\affiliation{Department of Physics, Humboldt-Universität zu Berlin, 10099 Berlin, Germany}

\author{Yaru Liu*}
\email{yaruliu879@jnu.edu.cn}
\affiliation{Potsdam Institute for Climate Impact Research, 14473 Potsdam, Germany}
\affiliation{Department of Mathematics, Jinan University, 510632 Guangzhou, China}

\author{Serhiy Yanchuk}
\affiliation{Potsdam Institute for Climate Impact Research, 14473 Potsdam, Germany}
\affiliation{Institute of Mathematics, Humboldt-Universität zu Berlin, 10099 Berlin, Germany}

\date{\today}

\begin{abstract}
Extreme multistability (EM) is characterized by the emergence of infinitely many coexisting attractors or continuous families of stable states in dynamical systems.  
EM implies complex and hardly predictable asymptotic dynamical behavior.
We analyse a model for pendulum clocks coupled by springs and suspended on an oscillating base, and show how EM can be induced in this system by a specifically designed coupling.
First, we uncover that symmetric coupling can increase the dynamical complexity.
In particular, the coexistence of multiple isolated attractors and continuous families of stable periodic states is generated in a symmetric cross-coupling scheme of four pendulums.
These coexisting infinitely many states are characterized by different levels of phase synchronization between the pendulums, including anti-phase and in-phase states. 
Some of the states are characterized by splitting of the pendulums into groups with silent sub-threshold and oscillating behavior, respectively.
The analysis of the basins of attraction further reveals the complex dependence of EM on initial conditions.
\end{abstract}

\maketitle

\begin{quotation}
The coexistence of several asymptotic stable states for a dynamical system with fixed parameter values is called multistability.
This phenomenon has been identified in diverse fields of science both experimentally and theoretically.
Which asymptotic state the system will converge to is determined solely by its initial state.
When the number of stable states is infinite, extreme multistability (EM) becomes a dominant feature.
Understanding EM and its control is an important issue, because systems with EM offer even greater flexibility than those with finite multistability when switching from one stable state to another.
We give an example of EM in a coupled pendulums model that takes into account an escapement mechanism as well as local and global couplings.
We have paid a particular attention to the coupling structure that leads to the emergence of EM.
\end{quotation}

\section{Introduction}\label{Sec:Introduction}
%
Complex networks have largely enriched our understanding of a variety of complex dynamical systems in many fields, such as biology, ecology, climatology, sociology, and others~\cite{Newman2018,Yanchuk2021a}.
By modeling real-world systems as networks in which collections of dynamic nodes are connected by static or adaptive edges, one can study collective behaviors both analytically and numerically~\cite{Blekhman1988,arenas2008synchronization,NEW03,gross2008adaptive,Cabral2022a}.

Synchronization is a ubiquitous dynamical phenomenon that has been observed in many natural and engineering systems~\cite{arenas2008synchronization,pikovsky_rosenblum_kurths_2001,osipov2007synchronization,boda2013kuramototype,chhabria2018targeted,ikeguchi2019analysis}.
Different types of synchronous patterns have been identified involving complete synchronization \cite{PEC98} (oscillators' states become asymptotically the same with time), cluster synchronization (a network splits into groups of synchronous elements)~\cite{dahms2012cluster,Lucken2012a}, special types of spatial coexistence of coherent  and incoherent states~\cite{abrams2004chimera,omelchenko2011loss,kasatkin2017selforganized}, and many others.
Various patterns have been found in experimental contexts, such as optoelectronic networks~\cite{soriano2013complex}, chemical networks~\cite{tinsley2012chimera}, neural networks~\cite{hammond2007pathological},  ecological~\cite{blasius1999complex}, and climate systems~\cite{tsonis2004architecture}.

Apart from synchronization, multistability -- the coexistence of several asymptotic stable states (attractors) for a given set of parameters -- is another intriguing phenomenon which has been studied for decades in modern nonlinear science~\cite{feudel2008complex,pisarchik2014control,dudkowski2022multistabilityb}.
The final state of a system with multistability depends crucially on initial conditions.
Multistability has also been observed in many areas of science, such as nonlinear optics~\cite{arecchi1991rate},
neuroscience~\cite{foss1996multistability},
climate dynamics~\cite{power1993multiple},
laser physics~\cite{masoller2002noiseinduced}, electronic oscillators~\cite{borresen2002further},
and in different classes of systems,
such as weakly dissipative systems~\cite{feudel1996map}, systems with time delays~\cite{Yanchuk2010a,balanov2005delayed}, and coupled systems~\cite{feudel1998dynamical}.

Understanding the emergence of co-existing attractors is an important issue, and controlling multistability is an even more difficult task.
When the number of co-existing attractors increases infinitely, EM emerges.
In coupled systems, the presence of EM has been found to be closely related to partial synchrony~\cite{hens2015extreme}.
By designing a specific coupling scheme to achieve partial synchrony, one can obtain infinitely many coexisting stable states~\cite{hens2015extreme,sun1999uncertain,ngonghala2011extremea,hens2012how,pal2014multistablea}.
Apart from the conservative cases, a common reason for the occurrence of EM in networks is time-reversibility, a special type of spatio-temporal symmetry \cite{Politi1986,Pikovsky-Rosenau,Lamb1998,ASH16a,Burylko2018}.

Despite the extensive literature on multistable dynamical systems, the emergence of multistability or EM in networked dynamical systems remains a challenging problem due to a large number of possible routes to EM, some of which have yet to be discovered.
Analytical and numerical challenges arise from the diversity of coupling topologies and the complexity of individual models.

In this work, we address the multistability problem in a mathematical model of coupled clocks suspended on a rotating disc and additionally coupled with springs.
The interaction of the clocks with the disc provides the global coupling among all clocks and therefore influences their behavior, similar to the interaction of the pedestrians with the bridge in the famous effect of crowd synchrony on the Millennium Bridge~\cite{Strogatz2005}. 
Such a global scheme has also proved useful in uncovering complex transient states~\cite{dudkowski2020transient}.
The oscillating clocks are also locally coupled via springs.
In Ref.~\onlinecite{dudkowski2020small}, a similar system of three coupled clocks was studied.

The following main results are obtained in this work:
\begin{itemize}
  \item We generalize the system of three coupled clocks \cite{dudkowski2020small} into a network-coupled scenario allowing arbitrary coupling configurations.
  \item We investigate how different coupling topologies affect the multistability in systems of three and four coupled clocks.
  We observe that more symmetric coupling topologies can lead to more complex dynamics with higher multistability.
  
  A particularly reach appears to be the ``cross-coupling'' structure with ``diagonal'' spring couplings in the system of four coupled clocks. In such a case, we observe EM that combines continuous family of stable attractors with different phase relations between the clocks. We provide an analytical and numerical description of this new phenomenon.
  \item
  Furthermore, we discuss how the discontinuity of the escapement mechanism affects the multistability in the system.
  The clocks within certain coupled groups (clusters) remain either silent or oscillating and in-phase synchronized due to the switching of the escapement mechanism. This leads to three qualitatively different discontinuity-induced types of attractors.
\end{itemize}

\begin{table*}[t]
    \caption{Parameters for the $N$-pendulums of system (\ref{Eq:general_model}).}\label{Tab:Symbol}
    \setlength{\tabcolsep}{10pt}
    \begin{tabularx}{\textwidth}{llX}
        \hline
        Parameter & Value & Definition  \\
        \hline
        $B_0$ & 5.115\ [kgm$^2$] & Support's moment of inertia \\
        $k_\varphi$ & 17.75\ [N/m] & Stiffness coefficient of springs \\
        $k_\theta$ & 34\ [Nm] & Stiffness of the spring connecting the base
        and the unmoving support\\
        $c_\theta$ & $\Delta=ln(2)$ &Damping of the supporting base \\
        $c_\varphi$ & 0.01\ [Nms] & Damping
        of the damper connecting the base and the unmoving
        support \\
        $m$ & 1\ [kg] & Mass of each pendulum clock\\
        $l$ & 0.24849\ [m] &  Lengths of the pendulums\\
        $g$ & 9.81\ [m/s$^2$] & Gravity acceleration\\
        $d$ & 1\ [m] & Distance between $O$ and $S_i$ $(i=1,2,...,N)$\\
        {$\alpha_i=\sphericalangle ({Ox},~\overline{OS_i})$}& $(\frac{360}{N})^\circ=\frac{2\pi}{N}$ & $Ox$ denotes the positive $x$ half axis\\
        $M$ & 0.075\ [Nm] & Fixed external momentum\\
        $\varepsilon_0$ & $5^\circ=5*\pi/180$ & Escapement threshold\\
        \hline
    \end{tabularx}
\end{table*}

\section{Model and measures}\label{Sec:Model and measures}
\subsection{General model}\label{Subsec:General model description}
\begin{figure*}[!htbp] \centering
\includegraphics[width=0.95\textwidth]{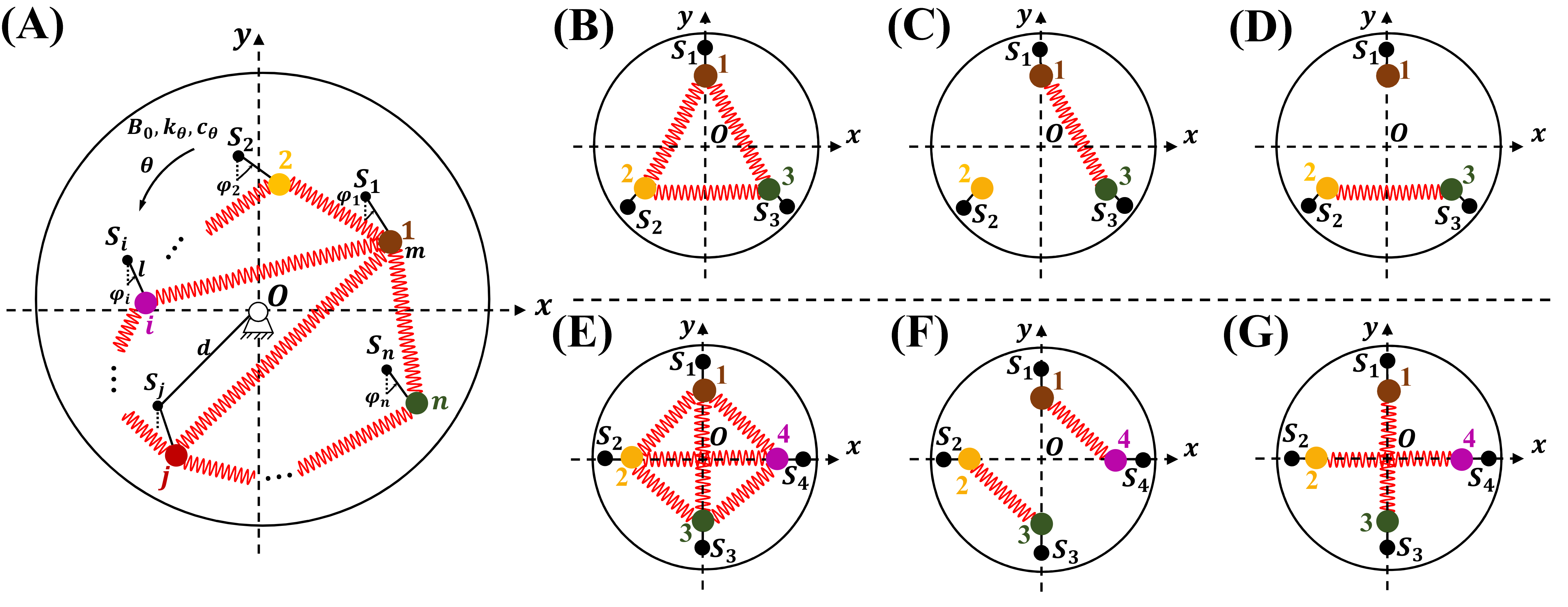}
    \caption{
    (A) The scheme of $N$ coupled identical pendulum clocks (shown as circles with different colors) suspended at evenly distributed black points $S_i$, $i = 1, 2,...,N$, on an oscillating supporting base (the $xy$ plane).
    The local coupling is realized using springs between the clocks.
    (B)--(D): For $N=3$, three types of coupling structures of springs include all-to-all, asymmetric, and symmetric topologies.
    (E)--(G): For $N=4$, three types of coupling structures of springs include all-to-all, asymmetric, and symmetric topologies.
    }
    \label{Fig:Model}
\end{figure*}

We first present a mathematical model of the $N$ coupled pendulum clocks suspended on a rotating disc, see~Fig.~\ref{Fig:Model}(A).
The rotating disc provides a global coupling, while the springs allow for arbitrary local coupling structure.
Our model is a generalization of the system of three pendulums from Ref.~\onlinecite{dudkowski2020small}.

The supporting base is placed at the origin $O$ of the $xy$ plane and it can oscillate freely around the axis perpendicular to the plane of Fig.~\ref{Fig:Model}; the angular deviation of the base is $\theta$.
The properties of the base are described by the moment of inertia $B_0$ [kgm$^2$], the stiffness of the spring connecting the base and the static support $k_\theta$ [Nm], and the damping $c_\theta$ [Nms]. 
Identical pendulums (marked colored filled circles) are suspended at evenly distributed black points $S_i$, $i = 1, 2, 3,...,N$, i.e., the angles between ${OS}_i$ and ${OS}_{i+1}$ (index $i$ is considered mod $N$) are ${{2\pi}/{N}}$.
The angles $\alpha_i=\sphericalangle({Ox},~\overline{OS}_{i})$ characterize the angular position of the suspension points $S_i$, where $Ox$ is the positive $x$ half-axis.
The parameter $d=|\overline{OS}_i|$ $(i=1,2,...,n)$ is the distance between the origin $O$ and each suspension point $S_i$.
Each pendulum is described by the angle displacement $\varphi_i$, the mass $m$ [kg], the length $l$ [m], and the damping coefficient $c_\varphi$ [Nms].
The stiffness coefficients of the springs are $k_\varphi$ [N/m].
The description and the values for all parameters are summarized in Table~\ref{Tab:Symbol}.

The equations of motion of the $N$ coupled pendulums is given by the following system:
\begin{widetext}
    \begin{equation}\label{Eq:general_model}
    \begin{array}{ll}
    \displaystyle
    (B_0+n mr^2)\ddot{\theta}+k_\theta\theta+c_\theta\dot{\theta}+
    \sum\limits_{i=1}^nmr\{l[\ddot{\varphi}_i \sin(\varphi_i-\theta-\alpha_i)+\dot{\varphi}^2_i \cos(\varphi_i-\theta-\alpha_i)]+g\cos(\alpha_i+\theta)\}+\Delta V_{\theta}=0,\\
    ml^2\ddot{\varphi}_i+ mgl\sin{\varphi_i}+c_ \varphi{\dot{\varphi}_i}+mrl[\ddot{\theta} \sin(\varphi_i-\theta-\alpha_i)-
    \dot{\theta}^2\cos(\varphi_i-\theta-\alpha_i)]+\Delta V_{\varphi_i} =M_{E_i}, \\
    \end{array}
    \end{equation}
\end{widetext}
where $i=1,2,...,N$.

The build-in escapement mechanism produces the moment of force, which is modeled by the discontinuous functions $M_{E_i}$, $i = 1, 2,3,...,N$ \cite{czolczynski2011why,kapitaniak2012synchronization}.
These functions depend not only on the displacement 
$\varphi_i(t)$, but also on the position of the $i$-th mechanism’s cogwheel versus the mechanism’s pallet $\sigma_i(t)$:
\begin{equation}\label{Eq:escapement_mechanism}
M_{E_i}=
\begin{cases}
M&: \sigma_i=1\wedge 0<\varphi_i<\varepsilon_0,\\
-M&: \sigma_i=2\wedge -\varepsilon_0<\varphi_i<0,\\
0&: \mathrm{otherwise.}
\end{cases}
\end{equation}
Here $M=0.075$ [Nm] represents the value of the external momentum, while $\varepsilon_0 = 5.0^\circ$ denotes the escapement's threshold (the mechanism turns off as the pendulum exceeds this threshold).
In fact, $\sigma_i(t)$ become additional discrete-valued variables in the system that are influencing the system's dynamics via the terms $M_{E_i}$ and which are changing discontinuously according to the following rules:
\begin{itemize}
    \item[(I)] When a pendulum $\varphi_i$ crosses the escapement threshold at some time moment $t^*$: $\varphi_i(t_*) = \varepsilon_0$ with increasing $\varphi_i$, i.e., $\dot \varphi_i(t_*)>0$, the variable $\sigma_i(t)$ is set to 2 for all $t\in [t^*,t_\mathrm{e})$, where $t_\mathrm{e}$ is the time of a next event.
    \item[(II)] When the pendulum $\varphi_i$ crosses the escapement threshold $\varphi_i(t_*) = - \varepsilon_0$ with decreasing $\varphi_i$, i.e., $\dot \varphi_i(t_*)<0$, the variable $\sigma_i(t)$ is set to 1 for all $t\in [t^*,~t_\mathrm{e})$, where $t_\mathrm{e}$ is the time moment of a next crossing event.
\end{itemize}
In this way, the variables $\sigma_i(t)$ are piece-wise constant with the possible discrete values 1 or 2, which change discontinuously when either event (I) or (II) occurs.

The terms $\Delta V_{\varphi_i}$ and $\Delta V_{\theta}$ in model (\ref{Eq:general_model}) describe the moments of forces from the coupling springs.
These two terms can be written explicitly using the following terms: $s_{ij}$, the constant distance between the $i$-th and $j$-th clocks when the system stays still,
and $\hat{s}_{ij}(t)$, the time-dependent distance between the $i$-th and $j$-th clocks for the moving system:
\begin{widetext}
\begin{equation}\label{Eq:springs_moments_of_forces_and_distance}
    \begin{array}{ll}
    \displaystyle
    s_{ij}=r\sqrt{2(1-\cos(\alpha_i-\alpha_j))},\\
    \hat{s}_{ij}=\sqrt{s_{ij}^2+2l^2(1-\cos(\varphi_i-\varphi_j))+8lr
    \sin\left(\frac{\varphi_i-\varphi_j}{2}\right)
    \sin\left(\frac{\alpha_i-\alpha_j}{2}\right)
    \sin\left(\frac{\varphi_i+\varphi_j-\alpha_i-\alpha_j}{2}-\theta\right)},\\
    \Delta V_{\theta}=2lrk_\varphi\sum\limits_{i=1}^n\sum\limits_{j=1}^na_{ij}
    \left(1-\frac{s_{ij}}{\hat{s}_{ij}}\right)
    \sin\left(\frac{\varphi_i-\varphi_j}{2}\right)\sin\left(\frac{\alpha_j-\alpha_i}{2}\right)\cos\left(\frac{\varphi_i+\varphi_j-\alpha_i-\alpha_j}{2}-\theta\right),\\
    \Delta V_{\varphi_i}=\sum\limits_{i=1}^na_{ij}k_\varphi l\left(1-\frac{s_{ij}}{\hat{s}_{ij}}\right)
    \left[l\sin(\varphi_i-\varphi_j)+2r
    \sin \left(\frac{\alpha_i-\alpha_j}{2} \right)
    \sin\left(\varphi_i-\frac{\alpha_i+\alpha_j}{2}-\theta\right)
    \right],
    \end{array}
\end{equation}
\end{widetext}
where $(a_{ij})$ is the coupling matrix via the springs, i.e., $a_{ij}=1$ if the pendulum $i$ is connected with the pendulum $j$ via a spring and $a_{ij}=0$ otherwise. $a_{ii}=0$ since there are no self-loops.

\begin{figure*}[!htbp]
\centering
\includegraphics[width=0.95\textwidth]{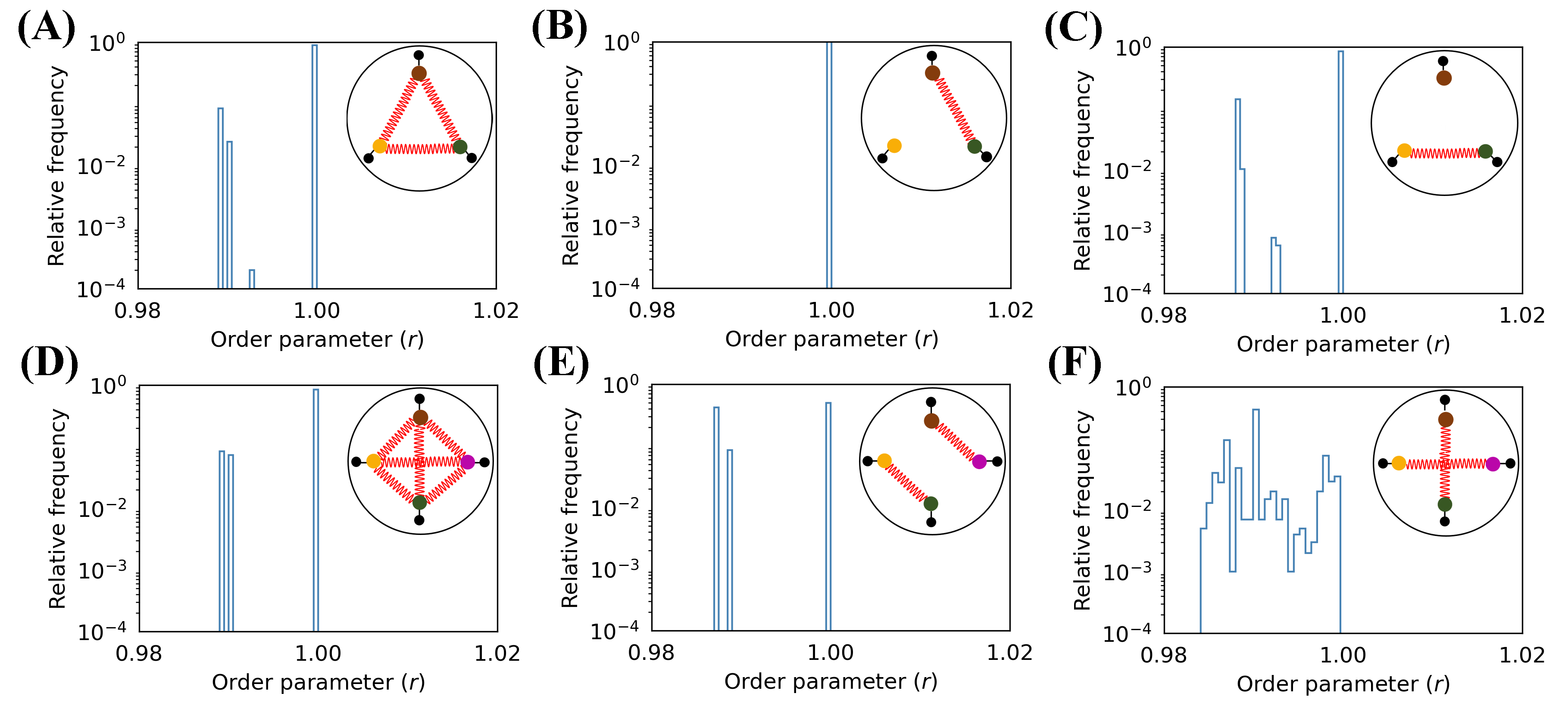}
\caption{
Distributions of the order parameter $r$ for different coupling topologies.
(A)-(C) show for 3-coupled pendulums ($N=3$), the identified multistabilities for all-to-all, asymmetric, and symmetric coupling structures (see right-upper corners), respectively.
For each coupling topology, 5000 order parameters are obtained from 5000 simulations using random initial conditions.
Parameters for $N=3$ are fixed as in Table \ref{Tab:Symbol}, in particular, $\alpha_1 = \frac{\pi}{2},$ $\alpha_2 = \frac{7\pi}{12}$, and $\alpha_3=\frac{11\pi}{12}$.
Each vertical bar corresponds to a potential attractor.
Similarly, (D)-(F) are multistabilities for 4-coupled pendulums ($N=4$), based on 1000 order parameters obtained from 1000 simulations.
Parameters for $N=4$ are also fixed as in Table \ref{Tab:Symbol}, in particular, $\alpha_1 = \frac{\pi}{2}$, $\alpha_2 = \pi$, $\alpha_3=\frac{3\pi}{2}$ and $\alpha_4=2\pi$.
For $N=3$, compared with (A) and (C), the asymmetric coupling structure in (B) decreases the dynamical complexity; while for $N=4$, compared with (E), (F) shows that the symmetric coupling can increase the dynamical complexity due to the emergence of EM.
}
\label{Fig:Coupling_R}
\end{figure*}

The influence of different coupling structures on collective dynamics has not been systematically reported for this model.
In the remaining part of this paper, we consider the cases $N=3$ (Figs.~\ref{Fig:Model}(B)-(D)) and $N=4$ (Figs.~\ref{Fig:Model}(E)-(G)). 
In particular, we focus on the following questions:
\begin{itemize}
  \item[\textbf{(1)}] How do different coupling topologies alter synchronization states and their basins of attraction?
  \item[\textbf{(2)}] Is there extreme multistability in the coupled pendulum model? If so, what is its origin?
\item[\textbf{(3)}] What are the effects of the discontinuity of the escapement mechanism on the dynamics and multistability?
\end{itemize}
We fix the parameters as in Table~\ref{Tab:Symbol}.
For $N=3$, the three considered types of coupling structures of springs include all-to-all (Fig.~\ref{Fig:Model}(B)), asymmetric (Fig.~\ref{Fig:Model}(C)), and symmetric (Fig.~\ref{Fig:Model}(D); mirror symmetry with respect to the vertical axis) topologies. 
Identical pendulums are suspended at evenly distributed points $S_i$, $i = 1, 2, 3$ with $\sphericalangle(\overline{OS}_i,~\overline{OS}_{i+1})=120^\circ$.
For $N=4$, we also consider three types of coupling structures of springs: all-to-all (Fig.~\ref{Fig:Model}(D)), asymmetric (Fig.~\ref{Fig:Model}(E)), and symmetric (Fig.~\ref{Fig:Model}(F)) topologies.
Here also the identical pendulums are suspended at evenly distributed points $S_i$, $i = 1, 2, 3, 4$ with $\sphericalangle(\overline{OS}_i,~\overline{OS}_{i+1})=90^\circ$.

We use Monte Carlo sampling and two classical measures (order parameter and mean frequencies) for the numerical quantification of synchronization states and the analysis of their basins of attraction.

\subsection{Measures}\label{Subsec:Measures}
\textit{Order parameter.}
We visualize the dynamics of the synchronization transitions with the Kuramoto order parameter:
\begin{equation}\label{Eq:order_parameter}
R(t)=\frac{1}{N}\sum\limits_{i=1} ^{N}e^{i\varphi_i(t)},
\end{equation}
where $N$ is the number of oscillators.
When $|R(t)|=1$ ($|R(t)|\approx0$), oscillators are in the complete synchronization (disordered) state.
The degree of synchronization in numerical simulations is quantified using the averaged value of the order parameter:
\begin{equation}\label{Eq:order_parameter_used}
r = \frac{1}{T_\mathrm{av}} 
\int_{T_\mathrm{tr}}^{T_\mathrm{tr}+T_\mathrm{av}}
|R(t)|dt.
\end{equation}
over the time interval $T_\mathrm{av}=50$ after a sufficiently long transient time $T_\mathrm{tr}$.

\textit{Mean frequency.}
The mean oscillation frequency of a pendulum is given as:
\begin{equation}\label{Eq:mean_frequency_used}
\langle\omega_i\rangle=\frac{2\pi n_i}{T_\mathrm{av}},
\end{equation}
where the same time interval of $T_\mathrm{av}=50$ is applied and $n_i$ represents the number of complete oscillations of $i$th clock within this interval.
The number of complete oscillations can be computed using the number of crossings of the Poincare map $\varphi_i=0$ or $\varphi_i=\varepsilon_0$.
The mean frequency is calculated using the last $T_\mathrm{av}=50$ time units after a sufficiently long transient time $T_\mathrm{tr}$.

\section{Collective dynamics for different coupling topologies}\label{Sec:Collective dynamics for different coupling topologies}
We first conduct various simulations of the system for random initial conditions. More specifically, we choose the following initial conditions $[\theta^0=0.01, \phi_1^0, \dots, \phi_n^0,  \dot{\theta^0}=0, \dot{\phi_1^0}=0, \dots, \dot{\phi_n^0}=0]$, where $\phi_1^0,\dots,\phi_n^0$, are chosen randomly from the interval $[-\pi, \pi)$. For the case $N=3$, simulations with 5,000 different  initial conditions are performed with the integration time 15,000, and the last 50 time units are used for the calculation of the order parameter and the mean frequency. 
For $N=4$, we perform 1,000 runs with the integration interval 10,000, and the transient 9,950.
The $r$ and $\langle\omega_i\rangle$ from Eqs. (\ref{Eq:order_parameter_used}) and (\ref{Eq:mean_frequency_used}), respectively, are used to estimate the synchronization state (attractor) in each simulation.
We found that further increase of the number of runs and the integration interval does not affect the obtained results qualitatively.

Figure~\ref{Fig:Coupling_R} shows the distribution of the order parameter $r$ for different initial conditions. 
This distribution reveals the possible number of different attractors. 
Figures~\ref{Fig:Coupling_R}(A)-(C) correspond to the coupling structures of three clocks in the Figs.~\ref{Fig:Model}(B)-(D), respectively. 
Here we see finitely many isolated lines indicating a relatively small  number of possible synchronization states. 
Interestingly, the case of asymmetric coupling in Fig.~\ref{Fig:Model}(C) exhibits lower dynamical complexity as that shown in Fig.~\ref{Fig:Coupling_R}(B), since only one line of the distribution of $r$ is achieved for all initial conditions.
Figures~\ref{Fig:Coupling_R}(A) and \ref{Fig:Coupling_R}(C) imply finite multistability with different possible asymptotic values of $r$.

For four coupled clocks (Figs.~\ref{Fig:Coupling_R}(D)-(F)), we also uncover that different coupling topologies lead to diverse dynamical complexities.
Specifically, three different lines of $r$ are observed in Figs.~\ref{Fig:Coupling_R}(D) and \ref{Fig:Coupling_R}(E).
More importantly, if the structure of the coupled clocks maintains the symmetry as Fig.~\ref{Fig:Model}(G), the distribution of the asymptotic order parameters in Fig. \ref{Fig:Coupling_R}(F) is no longer discrete, but contains continuous parts. 
Such a distribution indicates higher complexity and even EM. 
In order to characterize deeper the emergence of EM, we focus on analytical and numerical explanations of this phenomenon in the following sections.

\section{Extreme multistability}\label{Sec:Extreme multistability}
We recall that EM is potentially observed for the cross-coupling structure of four coupled clocks (Fig.~\ref{Fig:Model}(G)), where the distribution of asymptotic order parameters seems to be continuous  (Fig.~\ref{Fig:Coupling_R}(F)).
In this scheme, the opposite pendulums are connected by springs.
The corresponding coupling matrix has four nonzero entries $a_{13}=a_{31}=a_{24}=a_{42}=1$, and the angle position parameters are $\alpha_1 = \frac{\pi}{2}$, $\alpha_2 = \pi$, $\alpha_3=\frac{3\pi}{2}$ and $\alpha_4=2\pi$.
The system (\ref{Eq:general_model}) becomes:
\begin{widetext}
    \begin{eqnarray}\label{Eq:general_model_4}
    \begin{array}{ll}
    \displaystyle
    (B_0+4mr^2)\ddot{\theta}+k_\theta\theta+c_\theta\dot{\theta}
    +mrl[-\ddot{\varphi}_1 \cos(\theta-\varphi_1)-\dot{\varphi}^2_1 \sin(\theta-\varphi_1)]
    +mrl[\ddot{\varphi}_2 \sin(\theta-\varphi_2)-\dot{\varphi}^2_2 \cos(\theta-\varphi_2)] \\
    +mrl[\ddot{\varphi}_3 \cos(\theta-\varphi_3)+\dot{\varphi}^2_3 \sin(\theta-\varphi_3)]
    +mrl[-\ddot{\varphi}_4 \sin(\theta-\varphi_4)+\dot{\varphi}^2_4 \cos(\theta-\varphi_4)]
    +\Delta V_{\theta}=0,\\
    ml^2\ddot{\varphi}_1+ mgl\sin{\varphi_1}+c_ \varphi{\dot{\varphi}_1}+mrl[-\ddot{\theta} 
    \cos(\theta-\varphi_1)+
    \dot{\theta}^2\sin(\theta-\varphi_1)]+\Delta V_{\varphi_1} =M_{E_1}, \\
    ml^2\ddot{\varphi}_2+ mgl\sin{\varphi_2}+c_ \varphi{\dot{\varphi}_2}+mrl[\ddot{\theta} \sin(\theta-\varphi_2)+
    \dot{\theta}^2\cos(\theta-\varphi_2)]+\Delta V_{\varphi_2} =M_{E_2}, \\
    ml^2\ddot{\varphi}_3+ mgl\sin{\varphi_3}+c_ \varphi{\dot{\varphi}_3}+mrl[\ddot{\theta} \cos(\theta-\varphi_3)-
    \dot{\theta}^2\sin(\theta-\varphi_3)]+\Delta V_{\varphi_3} =M_{E_3}, \\
    ml^2\ddot{\varphi}_4+ mgl\sin{\varphi_4}+c_ \varphi{\dot{\varphi}_4}+mrl[-\ddot{\theta}\sin(\theta-\varphi_4)-
    \dot{\theta}^2\cos(\theta-\varphi_4)]+\Delta V_{\varphi_4} =M_{E_4}, \\
    \end{array}
    \end{eqnarray}
where
    \begin{equation}\label{Eq:springs_moments_of_forces_4}
    \begin{array}{ll}
    \displaystyle
    \Delta V_{\theta}=4lrk_\varphi \left[
    \left(1-\frac{s_{24}}{\hat{s}_{24}}\right)\sin\left(\frac{\varphi_2-\varphi_4}{2}\right)\sin\left(\theta-\frac{\varphi_2+\varphi_4}{2}\right)
    - \left(1-\frac{s_{13}}{\hat{s}_{13}}\right)\sin\left(\frac{\varphi_1-\varphi_3}{2}\right)\cos\left(\theta-\frac{\varphi_1+\varphi_3}{2}\right)
    \right],\\
    \Delta V_{\varphi_1}=k_\varphi l \left(1 - \frac{2r}{\hat s_{13}}\right)
    [l\sin(\varphi_1-\varphi_3)-2r\sin(\theta-\varphi_1)],\\
    \Delta V_{\varphi_2}=k_\varphi l\left( 1-\frac{2r}{\hat s_{24}}\right)
    [l\sin(\varphi_2-\varphi_4)-2r\cos(\theta-\varphi_2)],\\
     \Delta V_{\varphi_3}=k_\varphi l\left(1-\frac{2r}{\hat s_{31}}\right)
     [l\sin(\varphi_3-\varphi_1)+2r\sin(\theta-\varphi_3)],\\
    \Delta V_{\varphi_4}=k_\varphi l\left(1-\frac{2r}{\hat s_{24}}\right)[l\sin(\varphi_4-\varphi_2)+2r\cos(\theta-\varphi_4)].
    \end{array}
    \end{equation}
The distances $s_{ij}$ and $\hat s_{ij}$ are:
\label{Eq:distance_4}
    \begin{equation}\label{4-8}
    \begin{array}{ll}
    \displaystyle
    s_{13}=s_{31}=s_{24}=s_{42}=2r,\\
    \hat{s}_{13} = \hat{s}_{31}=\sqrt{4r^2+2r^2(1-\cos(\varphi_1-\varphi_3))-8lr\sin\left(\frac{\varphi_1-\varphi_3}{2}\right)\sin\left(\theta-\frac{\varphi_1+\varphi_3}{2}\right)},\\
    \hat{s}_{24} = \hat{s}_{42}=\sqrt{4r^2+2r^2(1-\cos(\varphi_2-\varphi_4))-8lr\sin\left(\frac{\varphi_2-\varphi_4}{2}\right)\cos\left(\theta-\frac{\varphi_2+\varphi_4}{2}\right)}.\\
    \end{array}
    \end{equation}
\end{widetext}

As we will see later, the regime of EM is characterized by the emergence of two frequency synchronized clusters each containing two clocks. 
The following phase relations are observed for the synchronized clusters:  ``in-phase-in-phase" (II), ``in-phase-anti-phase" (IA), ``anti-phase-in-phase" (AI) and ``anti-phase-anti-phase" (AA). The exact meaning of these relations are given in Table \ref{Tab:phase possibility}.
For example, IA means that the clocks in the first cluster are in-phase and  anti-phase in the second cluster.
Additionally, due to the discontinuity induced by the escapement mechanism, the mixed states are observed, when one or both of the clusters are not oscillating.
This is possible due to the fact that the clocks do not cross periodically the escapement threshold and, hence, do not gain energy.
The following clusters are observed: "silent-in-phase" (SI), "in-phase-silent" (IS), and "silent-silent" (SS). 

\begin{figure*}[!htbp] \centering
\includegraphics[width=0.95\textwidth]{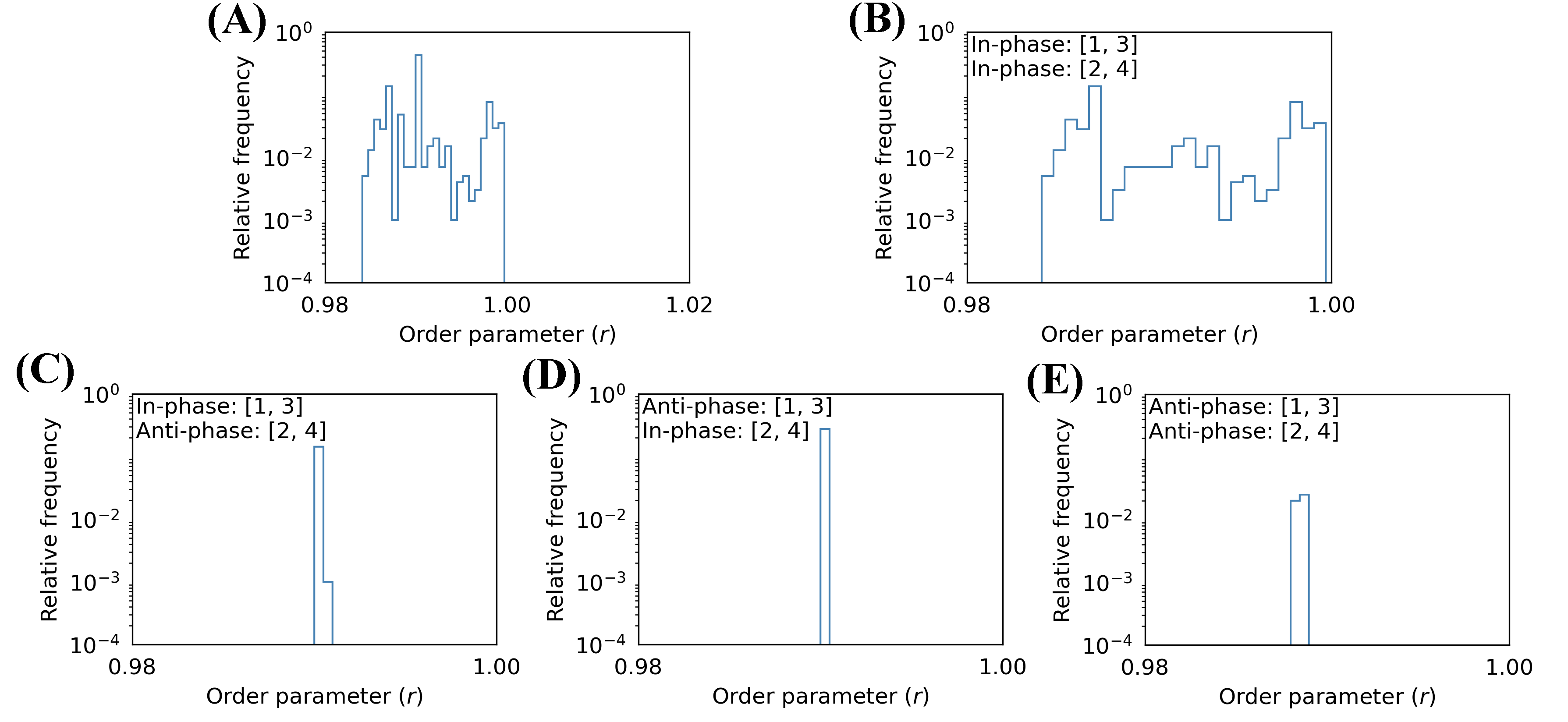}
\caption{
Distribution of the order parameter $r$ for 4 coupled clocks (\ref{Eq:general_model_4}) with the coupling topology as in Fig.~\ref{Fig:Model}(G). 
(A) is same as Fig. \ref{Fig:Coupling_R}(F), obtained by Monte Carlo sampling with 1,000 random trials. 
(B)-(E) represent the parts of the distribution (extracted from (A)) that correspond to specific cluster states: (B) counts only the order parameters for the trials ending in II (in-phase-in-phase) configuration, (C) stands for IA, (D) for AI, and (E) for AA, see Table \ref{Tab:phase possibility} explaining the cluster states. 
The main observation is that only case (B) is related to the emergence of EM.
Parameters are fixed as in Table \ref{Tab:Symbol} with $\alpha_1 = \frac{\pi}{2}$, $\alpha_2 = \pi$, $\alpha_3=\frac{3\pi}{2}$ and $\alpha_4=2\pi$ for the 4-coupled clocks ($N=4$).
}
\label{Fig:EM}
\end{figure*}

\begin{table}[!htbp]
    \caption{Different phase-clusters possibilities for 4-coupled clocks (\ref{Eq:general_model_4}) with cross-coupling structure (see Fig. \ref{Fig:Model}(G)).}
    \setlength{\tabcolsep}{2.5pt}
    \begin{tabularx}{0.48\textwidth}{ll}
    \hline 
    Symbol & Description \\
    \hline 
    II & In-phase-in-phase ($\varphi_1(t)=\varphi_3(t),\ \varphi_2(t)=\varphi_4(t)$)\\
    IA & In-phase-anti-phase ($\varphi_1(t)=\varphi_3(t), \ \varphi_2(t)=-\varphi_4(t)$)\\
    AI & Anti-phase-in-phase ($\varphi_1(t)=-\varphi_3(t), \ \varphi_2(t)=\varphi_4(t)$)\\
    AA & Anti-phase-anti-phase ($\varphi_1(t)=-\varphi_3(t), \ \varphi_2(t)=-\varphi_4(t)$)\\
    SI & Silent-in-phase ($\varphi_1(t)=\varphi_3(t)=0, \ \varphi_2(t)=\varphi_4(t)$)\\
    IS & In-phase-silent ($\varphi_1(t)=\varphi_3(t), \ \varphi_2(t)=\varphi_4(t)=0$)\\
    SS & Silent-silent ($\varphi_1(t)=\varphi_3(t)=0, \ \varphi_2(t)=\varphi_4(t)=0$)\\
    \hline 
    \end{tabularx}
    \label{Tab:phase possibility}
\end{table}

In Figure \ref{Fig:EM}, we split the probability distribution of the order parameter $r$ accordingly to the cluster states observed.  
Specifically, Fig.~\ref{Fig:EM}(A) gives the whole distributions, same as in Fig.~\ref{Fig:Coupling_R}(F)).
Figure~\ref{Fig:EM}(B) exacts from Fig.~\ref{Fig:EM}(A) only the order parameters that correspond to II phase clusters, Fig.~\ref{Fig:EM}(C) to IA, Fig.~\ref{Fig:EM}(D) to AI, and Fig.~\ref{Fig:EM}(E) to AA.
Only II clusters exhibit a continuous distribution of $r$, thus suggesting that EM appears due to such type of clusters. 
In contrast, Figs.~\ref{Fig:EM}(C)-(E) shows only a finite number of lines of $r$.
In the following sections, we provide additional analytical and numerical evidences that confirm our observation and explain the phenomenon of EM.

\subsection{Family of stable cluster states} \label{Subsec:Family of stable cluster regimes}
\subsubsection{Theoretical analysis of EM}
As Fig.~\ref{Fig:EM}(B) indicates, coexistence of infinitely many stable states can be related to the emergence of II, the in-phase-in-phase clusters.
To study the existence of such clusters, we show that the following subspace of II solutions:
\begin{equation}
\label{Eq:II_manifold}
\varphi_1(t)=\varphi_3(t)=\psi_1(t), \quad 
\varphi_2(t)=\varphi_4(t)=\psi_2(t),\quad \theta(t)=0.    
\end{equation}
is invariant with respect to the solutions of system (\ref{Eq:general_model_4}). 
Indeed, substituting $\varphi_1(t)=\varphi_3(t)=\psi_1(t)$ and $\varphi_2(t)=\varphi_4(t)=\psi_2(t)$ into (\ref{Eq:general_model_4}), we obtain $\Delta V_{\theta}=0$ and $\Delta V_{\varphi_i}=0$, and the equations for the new variables $\theta$, $\psi_1$, and $\psi_2$ read:
\begin{widetext}
    \begin{subequations}\label{Eq:general_model_4_EM}
    \begin{align}
    &(B_0+4mr^2)\ddot{\theta}+k_\theta\theta+c_\theta\dot{\theta}=0,\label{Eq:general_model_4_EM_0}\\
    &ml^2\ddot{\psi}_1+ mgl\sin{\psi_1}+c_ \varphi{\dot{\psi}_1}+mrl[-\ddot{\theta} 
    \cos(\theta-\psi_1)+
    \dot{\theta}^2\sin(\theta-\psi_1)]=M_{E_1},\label{Eq:general_model_4_EM_1}\\
    &ml^2\ddot{\psi}_2+ mgl\sin{\psi_2}+c_ \varphi{\dot{\psi}_2}+mrl[\ddot{\theta} \sin(\theta-\psi_2)+
    \dot{\theta}^2\cos(\theta-\psi_2)] =M_{E_2},\label{Eq:general_model_4_EM_2}\\
    &ml^2\ddot{\psi}_1+ mgl\sin{\psi_1}+c_ \varphi{\dot{\psi}_1}+mrl[\ddot{\theta} \cos(\theta-\psi_1)-
    \dot{\theta}^2\sin(\theta-\psi_1)]=M_{E_1},\label{Eq:general_model_4_EM_3}\\
    &ml^2\ddot{\psi}_2+ mgl\sin{\psi_2}+c_ \varphi{\dot{\bar{\varphi}}_2}+mrl[-\ddot{\theta}\sin(\theta-\psi_2)-
    \dot{\theta}^2\cos(\theta-\psi_2)]=M_{E_2},\label{Eq:general_model_4_EM_4}
    \end{align}
    \end{subequations}
\end{widetext}
with
\begin{equation}\label{Eq:escapement_mechanism_4_EM}
M_{E_i}=
\begin{cases}
M&: \sigma_i=1\wedge 0<\psi_i<\varepsilon_0\\
-M&: \sigma_i=2\wedge -\varepsilon_0<\psi_i<0\\
0&: \mathrm{otherwise}
\end{cases}
\end{equation}
where $i=1,2$, $M = 0.075$ [Nm], and $\varepsilon = 5.0^\circ$.

Now, by setting $\theta=0$, we observe that Eq. (\ref{Eq:general_model_4_EM_0}) is satisfied, and the dynamical equations for $\psi_1$ (Eqs. (\ref{Eq:general_model_4_EM_1}) and (\ref{Eq:general_model_4_EM_3})) and for $\psi_2$ (Eqs. (\ref{Eq:general_model_4_EM_2}) and (\ref{Eq:general_model_4_EM_4})) become the same:
\begin{subequations}\label{Eq:general_model_4_EM_final}
\begin{align}
&ml^2\ddot{\psi}_1+ mgl\sin{\psi_1}+c_ \varphi{\dot{\psi}_1}=M_{E_1}, \label{Eq:general_model_4_EM_final_1} \\
&ml^2\ddot{\psi}_2+ mgl\sin{\psi_2}+c_ \varphi{\dot{\psi}_2}=M_{E_2}. \label{Eq:general_model_4_EM_final_2}
\end{align}
\end{subequations}
Thus, we have proven the following:

\newtheorem{prop}{Proposition}

\begin{prop}[II-cluster subspace]\label{II-cluster subspace}
    The subspace (\ref{Eq:II_manifold}) of the cluster II solutions is invariant with respect to the solutions (flow) of system (\ref{Eq:general_model_4}).
    This subspace is 4-dimensional ($S^2\times \mathrm{R}^2$), and the flow on this subspace is given by Eqs.~(\ref{Eq:general_model_4_EM_final_1}) and (\ref{Eq:general_model_4_EM_final_2}) which describe the relative motion of the clusters.
\end{prop}

Another important observation is that the dynamical equations  (\ref{Eq:general_model_4_EM_final_1}) for $\psi_1$ and (\ref{Eq:general_model_4_EM_final_2}) $\psi_2$ in the II-cluster subspace are (i) the same, and (ii) \textit{uncoupled from each other}, i.e., the equation for $\psi_1$ is independent on $\psi_2$ and vise-versa.
The latter property leads to the coexistence of infinitely many asymptotic states (and to EM eventually).
We formulate the corresponding result as a proposition.

\begin{prop}[EM of II-cluster states]\label{EM of II-cluster states}
    Assume that $c_\varphi,m,l,g,k_\theta,c_\theta$ and $B_0$ are positive parameters.
    Assume also that system (\ref{Eq:general_model_4_EM_final_1}) (or, equivalently, (\ref{Eq:general_model_4_EM_final_2})) possesses a stable nontrivial asymptotic state (attractor), and $\psi^*(t)$ is a solution on this attractor (i.e., a single clock has a stable oscillatory state). 
    Then the system on the II-cluster subspace (\ref{Eq:general_model_4_EM_final_1})--(\ref{Eq:general_model_4_EM_final_2}) possesses the following asymptotic states:\\
    -- II clusters: 
    \begin{equation}
        \label{Eq:II}
        \varphi_1=\varphi_3=\psi_1=\psi^*(t), \quad
        \varphi_2=\varphi_4=\psi_2=\psi^*(t+\gamma).
    \end{equation}
    -- SI clusters: 
    \begin{equation}
        \label{Eq:SI}
        \varphi_1=\varphi_3=\psi_1 = 0,\quad 
        \varphi_2=\varphi_4=\psi_2 =\psi^*(t).
    \end{equation}
    -- SS clusters: 
    \begin{equation}
        \label{Eq:SS}
        \varphi_1=\varphi_3=\psi_1 =0,\quad
        \varphi_2=\varphi_4=\psi_2=0.
    \end{equation}
    -- IS clusters: 
    \begin{equation}
        \label{Eq:IS}
        \varphi_1=\varphi_3=\psi_1 =\psi^*(t),\quad
        \varphi_2=\varphi_4=\psi_2=0.
    \end{equation}
    where $\gamma$ is an arbitrary real constant describing a phase shift between the clusters.
    Moreover, if $\gamma^*(t)$ is an orbitally asymptotically stable limit cycle (stable periodic oscillations of the clock), then the states (\ref{Eq:II}) build a stable invariant torus foliated by (infinitely many) periodic solutions of the form (\ref{Eq:II}).
\end{prop}

The main message of Proposition \ref{EM of II-cluster states} is that under "normal conditions" when the single clock oscillates periodically, the coupled system can have stable cluster II oscillations with an arbitrary phase-shift between the clusters.
If the phase-shift is zero, the order parameter $r$ is highest and equal 1, while it can achieve a continuous range of smaller values depending on the phase shift. The coexistence of such states leads to EM.
Note that the states SI, SS, and IS do not lead to EM, but correspond to isolated attractors in the coupled system.

\textbf{Proof of Proposition \ref{EM of II-cluster states}.}
Let us first mention that the equilibrium $\psi_1=\psi_2=0$ is asymptotically stable, and it corresponds to the stability of a silent state of a pendulum with damping and without external energy inflow.
Therefore, the two independent systems (\ref{Eq:general_model_4_EM_final_1}) and (\ref{Eq:general_model_4_EM_final_2}) can reach both attractors: zero equilibrium and non-trivial attractor corresponding to $\psi^*(t)$, depending on initial conditions. 
Moreover, an arbitrary phase shift $\psi^*(t+\gamma)$ is clearly also possible and belong to the same nontrivial attractor.
This provides the existence of the states (\ref{Eq:II})--(\ref{Eq:IS}).

The invariant torus from the proposition corresponds to the direct product of the limit cycles $\mathcal{C}\times \mathcal{C}$, where $\mathcal{C}=\{(\psi,~\dot \psi)\in (S^1\times \mathbb{R}): \psi=\gamma^*(t),~t\in \mathbb{R}\}$.
The stability of this torus follows from the orbital stability of the limit cycle in each of the subsystem and the properties of the cross-product of the uncoupled system (\ref{Eq:general_model_4_EM_final_1})--(\ref{Eq:general_model_4_EM_final_2}). \textbf{End of proof.}

\begin{figure*}[!htbp] \centering
\includegraphics[width=0.95\textwidth]{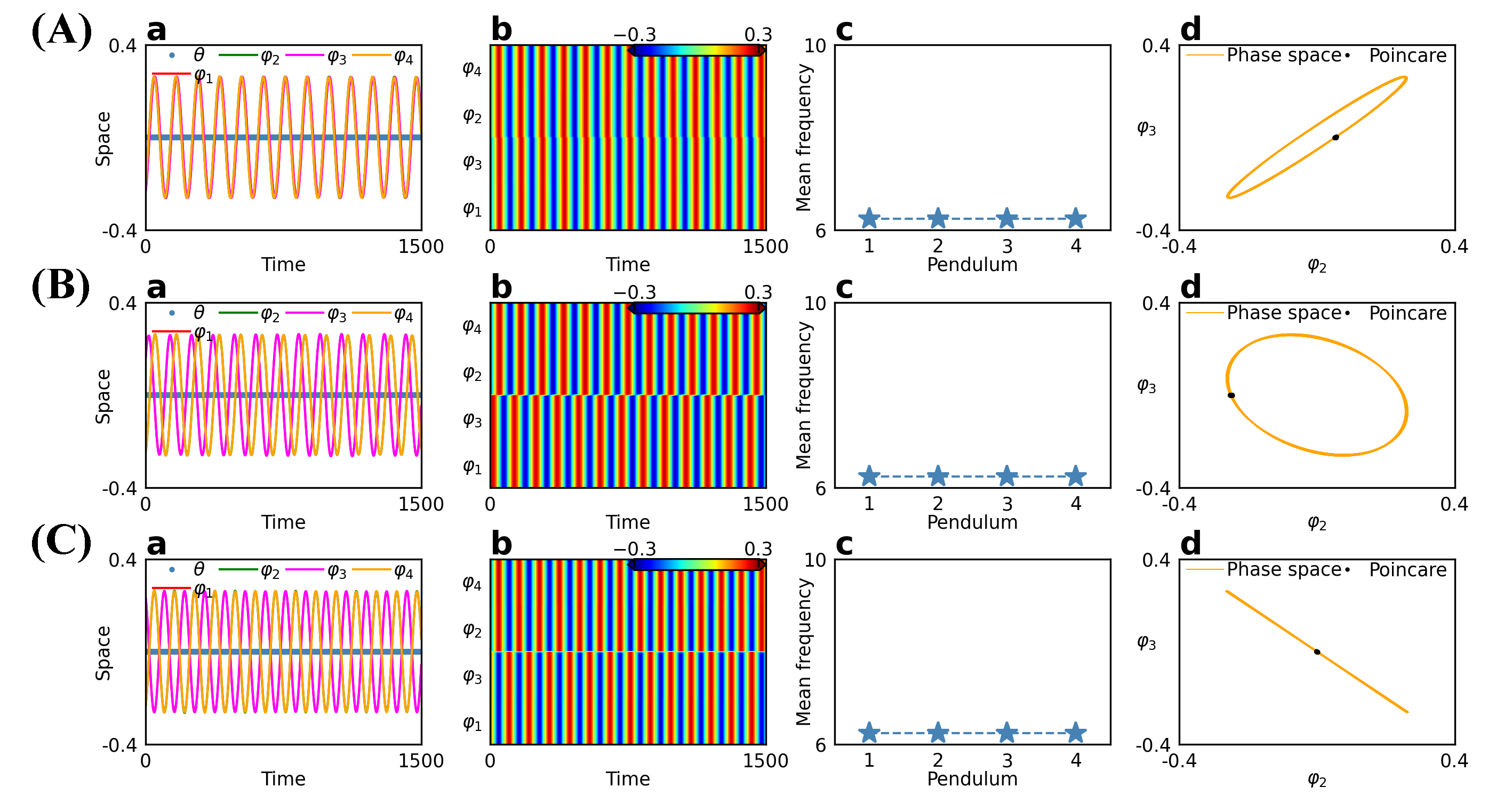}
\caption{
Three exemplary dynamical patterns for the regime of EM (see Fig. \ref{Fig:EM}(B)) for 4-coupled pendulums with cross-coupling structure (\ref{Eq:general_model_4}).
(A) Almost complete synchronization with a small phase shift between the clusters of clocks.
(B) Phase synchronization with an intermediate phase shift between the clusters.
(C) Anti-phase synchronization between the clusters.
Details of column figures are as follows:
{\bf{a}}: the time series for variables $\theta$, $\varphi_1$, $\varphi_2$, $\varphi_3$, $\varphi_4$, respectively,\ {\bf{b}}: the phase-time plots of the pendula,\ {\bf{c}}: the mean frequencies of the clocks,\ {\bf{d}}: Projections on two phase variables form different clusters (orange lines) and Poincaré maps (black dots).
Parameters are fixed as in Table \ref{Tab:Symbol} with $\alpha_1 = \frac{\pi}{2}$, $\alpha_2 = \pi$, $\alpha_3=\frac{3\pi}{2}$ and $\alpha_4=2\pi$ for the 4-coupled clocks ($N=4$).
}
\label{Fig:Dynamical_Patterns_Of_EM}
\end{figure*}

\begin{figure*}[!htbp] \centering
\includegraphics[width=0.95\textwidth]{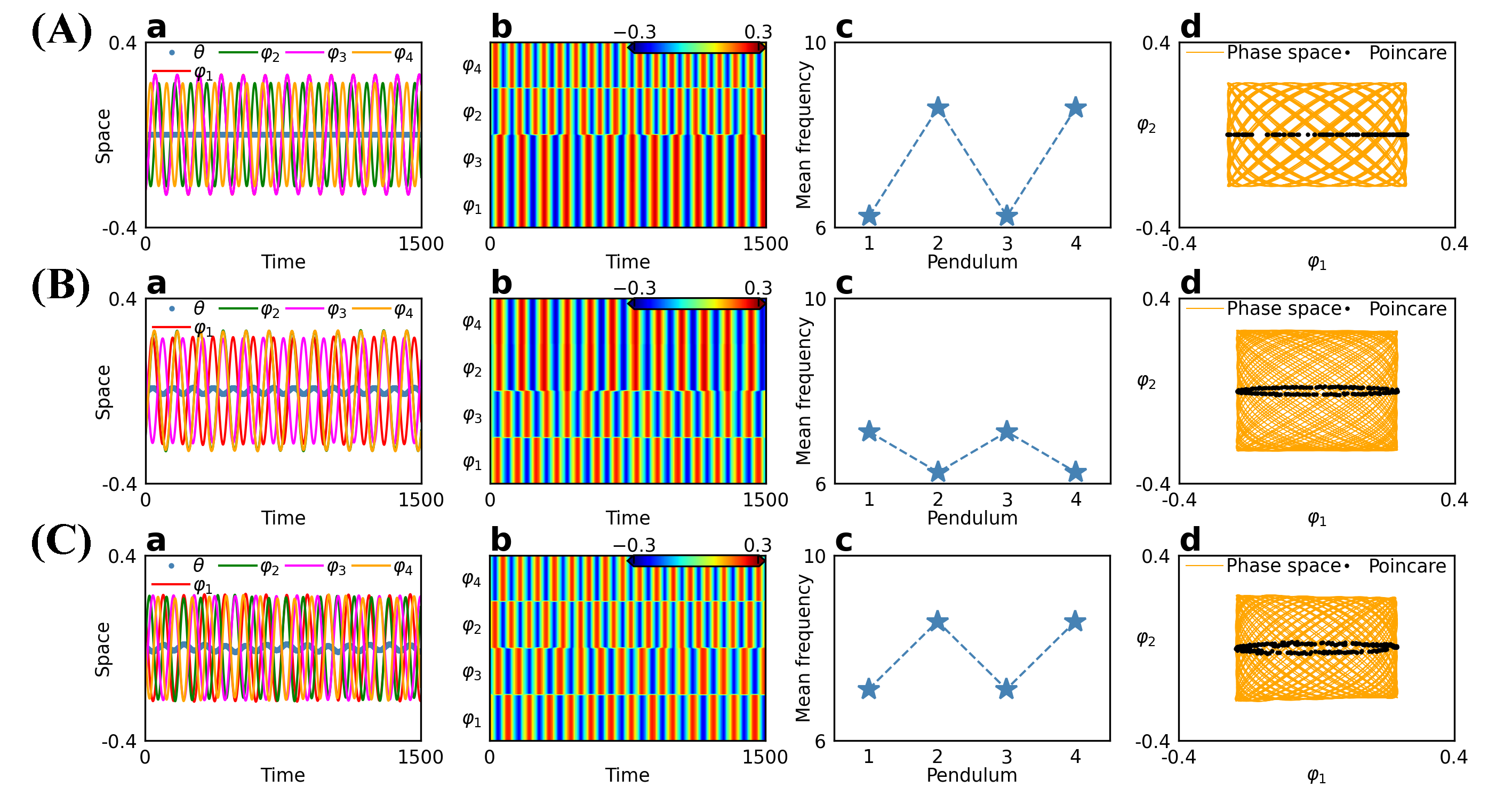}
\caption{
Dynamical patterns of three types of isolated attractors with large-amplitude oscillations for 4 coupled pendulums with cross-coupling structure (\ref{Eq:general_model_4}).
(A)-(C): IA, AI, and AA synchronization patterns (see Table~\ref{Tab:phase possibility} for the explanation of the abbreviations). IA, AI, and AA correspond to the order parameters from Figs.~\ref{Fig:EM}(C), (D), and (E), respectively.
Information on the columns:  
{\bf{a}}: the time series for variables $\theta$, $\varphi_1$, $\varphi_2$, $\varphi_3$, $\varphi_4$, respectively;\ {\bf{b}}: phase-time plots of the pendulum angles;\ {\bf{c}} mean frequencies;\ {\bf{d}}: projection of the solution on the phase variables from different clusters (orange lines) and Poincaré maps (black dots).
Parameters are fixed as in Table \ref{Tab:Symbol} with $\alpha_1 = \frac{\pi}{2}$, $\alpha_2 = \pi$, $\alpha_3=\frac{3\pi}{2}$ and $\alpha_4=2\pi$ for the 4-coupled clocks ($N=4$).
}
\label{Fig:Attractor_IA_AI_AA}
\end{figure*}

\subsubsection{Numerical study of EM}
Figure~\ref{Fig:Dynamical_Patterns_Of_EM} shows three examples of different stable II synchronization patterns from the continuous family of solutions by Eq. (\ref{Eq:II}).
Figure~\ref{Fig:Dynamical_Patterns_Of_EM} provides (A) almost in-phase, (B) a phase-shifted, and (C) anti-phase relations between the clusters. The dynamics within the clusters is completely synchronized: $\varphi_1=\varphi_3$ and $ \varphi_2=\varphi_4$.
These three states are exemplary, and different phase shifts are obtained from different initial conditions.
In spite of the phase shift, all pendulums are (mean) frequency synchronized (see the third column of Fig. \ref{Fig:Dynamical_Patterns_Of_EM}), since they follow the same motion according to Eq. (\ref{Eq:II}), only phase-shifted.
The fourth column of Fig.~\ref{Fig:Dynamical_Patterns_Of_EM} illustrates the phase-shift between the clusters and their periodic motion (orange curve). The black point shows the Poincaré map defined by $\varphi_1$ = 0 and $\dot{\varphi}_1 > 0$.

As for the order parameters $r$, Fig.~\ref{Fig:Dynamical_Patterns_Of_EM}(A) presents the trial leading to a relatively high order parameter close to 1 (almost complete synchronization) in Fig. \ref{Fig:EM}(B); while Fig.~\ref{Fig:Dynamical_Patterns_Of_EM}(C) represents the trial which falls into the left side (the inter-group anti-phase synchronization) of the order parameter distribution in Fig.~\ref{Fig:EM}(B).
The more trials one draws from random initial conditions, the more likely one can fill the gap regarding the order parameter between the inter-group anti-phase synchronization and complete synchronization to generate EM.

\begin{figure*}[!htbp] \centering
\includegraphics[width=0.95\textwidth]{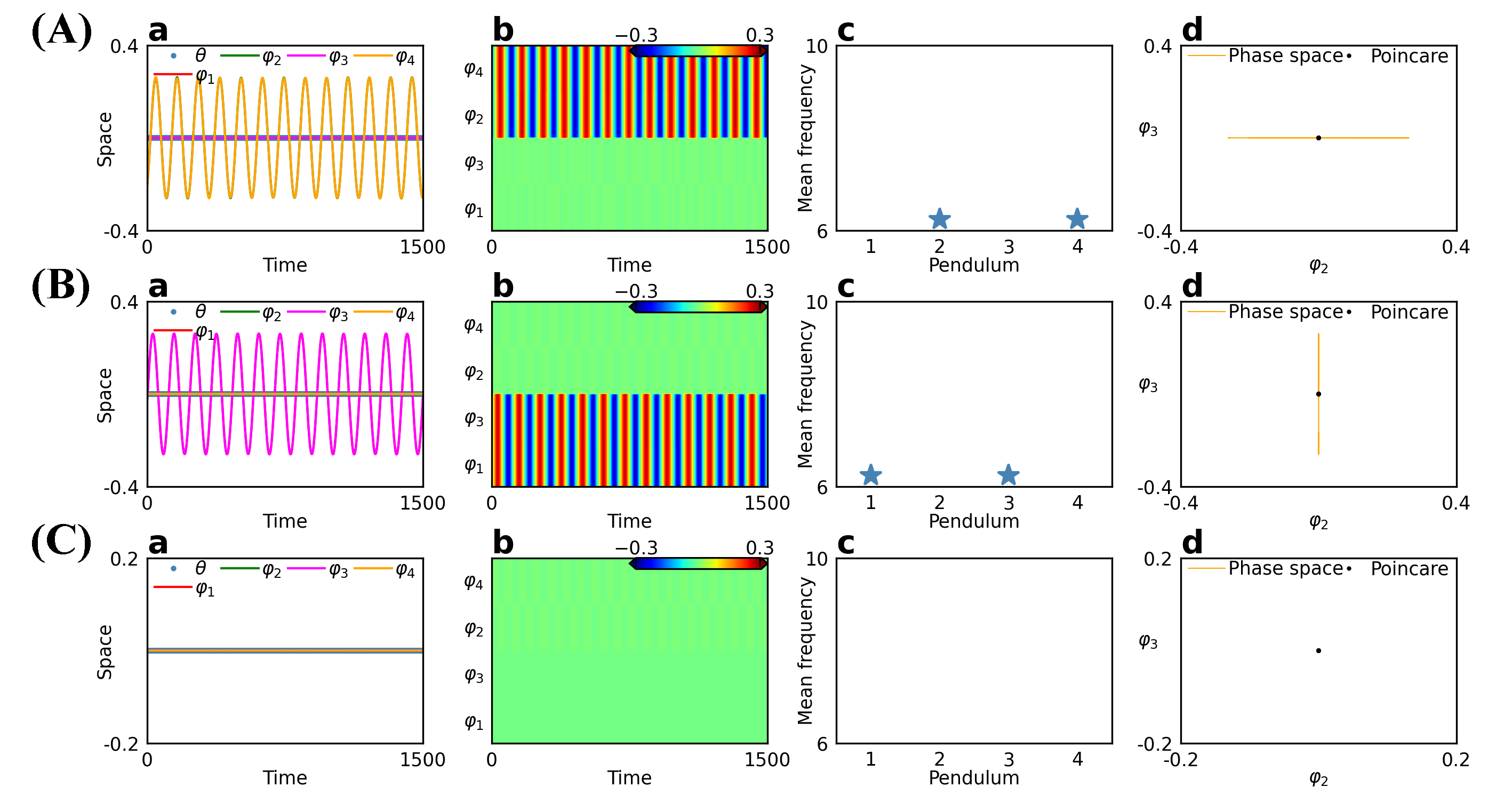}
\caption{
Dynamical patterns of three types of isolated attractors with small-amplitude oscillations for 4 coupled pendulums with cross-coupling structure (\ref{Eq:general_model_4}).
(A)-(C): SI, IS, and SS synchronization patterns respectively,  (see Table \ref{Tab:phase possibility} for the pattern explanations).
Information along the columns:
{\bf{a}}: time series for variables $\theta$, $\varphi_1$, $\varphi_2$, $\varphi_3$, and $\varphi_4$,\ {\bf{b}}: phase-time plots,\ {\bf{c}}:  mean frequencies (not available if a pendulum converges to 0),\ {\bf{d}}: projections on the $(\varphi_2,\varphi_3)$-plane (orange line) and  Poincaré maps (black dots, defined by $\varphi_2=0$ and $\dot{\varphi}_2 > 0$, $\varphi_1=0$ and $\dot{\varphi}_1 > 0$, and $\varphi_1=0$ and $\dot{\varphi}_1 > 0$, for (A), (B), and (C), respectively).
Parameters are fixed as in Table \ref{Tab:Symbol} with $\alpha_1 = \frac{\pi}{2}$, $\alpha_2 = \pi$, $\alpha_3=\frac{3\pi}{2}$ and $\alpha_4=2\pi$ for the 4-coupled clocks ($N=4$).
}
\label{Fig:Attractor_Discontinuity}
\end{figure*}

\subsection{Isolated attractors}\label{Subsec:Isolated attractors}
In addition to infinitely many stable II states from EM described above, system (\ref{Eq:general_model_4}) possesses coexisting isolated attractors corresponding to other synchronization patterns. 
These states can also be treated analytically and numerically in more detail.
However, since the main focus of this work is the EM phenomenon, we consider here exemplary only isolated attractors corresponding to IA patterns.

\subsubsection{Three types of isolated attractors with large-amplitude oscillations}
\paragraph*{Theoretical analysis of IA patterns.}
The IA solutions are characterized by the phase relations $\varphi_1(t)=\varphi_3(t)=\psi_1$ and $\varphi_2(t)=-\varphi_4(t)=\psi_2$.
Substituting this into system (\ref{Eq:general_model_4}), we obtain:
\begin{widetext}
    \begin{eqnarray}\label{Eq:general_model_4_IA}
    \begin{array}{ll}
    \displaystyle
    (B_0+4mr^2)\ddot{\theta}+k_\theta\theta+c_\theta\dot{\theta}+mrl[\ddot{\psi}_2 \sin(\theta-\psi_2)-\dot{\psi}^2_2 \cos(\theta-\psi_2)]
    +mrl[\ddot{\psi}_2 \sin(\theta+\psi_2)-\dot{\psi}^2_2 \cos(\theta+\psi_2)]
    +\Delta V_{\theta}=0,\\
    ml^2\ddot{\psi}_1+ mgl\sin{\psi_1}+c_ \varphi{\dot{\psi}_1}+mrl[-\ddot{\theta} 
    \cos(\theta-\psi_1)+
    \dot{\theta}^2\sin(\theta-\psi_1)]+\Delta V_{\varphi_1} =M_{E_1}, \\
    ml^2\ddot{\psi}_2+ mgl\sin{\psi_2}+c_ \varphi{\dot{\psi}_2}+mrl[\ddot{\theta} \sin(\theta-\psi_2)+
    \dot{\theta}^2\cos(\theta-\psi_2)]+\Delta V_{\varphi_2} =M_{E_2}, \\
    ml^2\ddot{\psi}_1+ mgl\sin{\psi_1}+c_ \varphi{\dot{\psi}_1}+mrl[\ddot{\theta} \cos(\theta-\psi_1)-
    \dot{\theta}^2\sin(\theta-\psi_1)]+\Delta V_{\varphi_3} =M_{E_3}, \\
    -ml^2\ddot{\psi}_2-mgl\sin{\psi_2}-c_ \varphi{\dot{\psi}_2}+mrl[-\ddot{\theta}\sin(\theta+\psi_2)-
    \dot{\theta}^2\cos(\theta+\psi_2)]+\Delta V_{\varphi_4} =M_{E_4}. \\
    \end{array}
    \end{eqnarray}
\end{widetext}
In the following, we introduce the new dimensionless parameter $\varepsilon=\frac{mrl}{B_0+4mr^2}$. For the chosen setup as in Table~\ref{Tab:Symbol}, we have $\varepsilon=0.02726$, i.e., $\varepsilon$ is small, and we will employ it in our analysis.
In fact, the smallness of this parameter is one of the reasons for the emergence of IA patterns.

Defining further $\bar{k}_\theta=\frac{k_\theta}{B_0+4mr^2}=3.7301$, $\bar{c}_\theta=\frac{c_\theta}{B_0+4mr^2}=0.02407$, $F(\theta,~\psi_2)=-[\ddot{\psi}_2 \sin(\theta-\psi_2)-\dot{\psi}^2_2 \cos(\theta-\psi_2)]
+[\ddot{\psi}_2 \sin(\theta+\psi_2)-\dot{\psi}^2_2 \cos(\theta+\psi_2)]
-\frac{\Delta V_{\theta}}{mrl}$ and $\theta=\varepsilon\psi$, the system (\ref{Eq:general_model_4_IA}) can be rewritten in the following form:
\begin{widetext}
    \begin{equation}\label{Eq:general_model_4_IA_perturbation}
    \begin{array}{ll}
    \displaystyle
    \ddot{\psi}+\bar{c}_\theta\dot{\psi}+\bar{k}_\theta\psi= F(\varepsilon\psi,~\psi_2),\\
    ml^2\ddot{\psi}_1+ mgl\sin{\psi_1}+c_ \varphi{\dot{\psi}_1}+\varepsilon mrl[-\ddot{\psi} 
        \cos(\varepsilon\psi-\psi_1)+\varepsilon\dot{\psi}^2\sin(\varepsilon\psi-\psi_1)]+\Delta V_{\varphi_1} =M_{E_1}, \\      
    ml^2\ddot{\psi}_2+ mgl\sin{\psi_2}+c_ \varphi{\dot{\psi}_2}+\varepsilon mrl[\ddot{\psi} \sin(\varepsilon\psi-\psi_2)+
        \varepsilon\dot{\psi}^2\cos(\varepsilon\psi-\psi_2)]+\Delta V_{\varphi_2} =M_{E_2},\\ 
    ml^2\ddot{\psi}_1+ mgl\sin{\psi_1}+c_ \varphi{\dot{\psi}_1}+\varepsilon mrl[\ddot{\psi}\cos(\varepsilon{\psi}-\psi_1)-
        \varepsilon\dot{\psi}^2\sin(\varepsilon{\psi}-\psi_1)]+\Delta V_{\varphi_3} =M_{E_3},\\
    -ml^2\ddot{\psi}_2-mgl\sin{\psi_2}-c_ \varphi{\dot{\psi}_2}+\varepsilon mrl[-\ddot{\psi}\sin(\varepsilon{\psi}+\psi_2)-\varepsilon \dot{\psi}^2\cos(\varepsilon{\psi}+\psi_2)]+\Delta V_{\varphi_4} =M_{E_4},
    \end{array}
    \end{equation}
where
    \begin{equation}\label{Eq:springs_moments_of_forces_4_IA}
    \begin{array}{ll}
    \displaystyle
        \Delta V_{\theta}=4lrk_\varphi
        \left(1-\frac{2r}{\sqrt{4r^2+2r^2(1-\cos(2\psi_2))-8lr\sin\psi_2\cos(\varepsilon\psi)}}\right)
        \sin\psi_2\sin(\varepsilon\psi),\\
        \Delta V_{\varphi_1}= \Delta V_{\varphi_3} =0,\\
        \Delta V_{\varphi_2}=k_\varphi l 
        \left(1-\frac{2r}{\hat{s}_{24}}\right)
        [l\sin(2\psi_2)-2r\cos(\varepsilon\psi-\psi_2)],\\
        \Delta V_{\varphi_4}=k_\varphi l
        \left(1-\frac{2r}{\hat{s}_{24}}\right)
        [-l\sin(2\psi_2)+2r\cos(\varepsilon\psi+\psi_2)],
    \end{array}
    \end{equation}
\end{widetext}
and the values for the distances satisfy:
\begin{equation}\label{Eq:distance_4_IA}
    \begin{array}{ll}
    \displaystyle
        s_{13}=s_{31}=s_{24}=s_{42}=2r,\\
        \hat{s}_{13}=\hat{s}_{31}=2r,\\
        \hat{s}_{24}=\hat{s}_{24}=\sqrt{4r^2+2r^2(1-\cos(2\psi_2))-8lr\sin\psi_2\cos(\varepsilon\psi)}.
    \end{array}
\end{equation}

In the zeroth-order in $\varepsilon$, system (\ref{Eq:general_model_4_IA_perturbation}) is reduced to
\begin{subequations}\label{Eq:general_model_4_IA_perturbation_0}
\begin{align}
&\ddot{\psi}+\bar{c}_\theta\dot{\psi}+\bar{k}_\theta\psi=F(0,~\psi_2),\label{eqn-11-1}\\
&ml^2\ddot{\psi}_1+ mgl\sin{\psi_1}+c_ \varphi{\dot{\psi}_1} =M_{E_1},\label{eqn-11-2}\\
&ml^2\ddot{\psi}_2+ mgl\sin{\psi_2}+c_ \varphi{\dot{\psi}_2}+\Delta V_{\varphi_2}=M_{E_2},\label{eqn-11-3}\\
&ml^2\ddot{\psi}_1+ mgl\sin{\psi_1}+c_ \varphi{\dot{\psi}_1} =M_{E_1},\label{eqn-11-4}\\
&-ml^2\ddot{\psi}_2-mgl\sin{\psi_2}-c_ \varphi{\dot{\psi}_2}-\Delta V_{\varphi_2} =-M_{E_2},\label{eqn-11-5}
\end{align}
\end{subequations}
from which one can see that Eq.~(\ref{eqn-11-2}) is equivalent to Eq.~(\ref{eqn-11-4}) and  Eq.~(\ref{eqn-11-3}) to Eq.~(\ref{eqn-11-5}). Hence, in this approximation, the subspace of the IA solutions:
\begin{equation}
\label{Eq:subspace_IA_solution}
\varphi_1(t)=\varphi_3(t)=\psi_1,\quad \varphi_2(t)=-\varphi_4(t)=\psi_2, \quad \theta(t)=0.
\end{equation}
is invariant.
For nonzero but small $\varepsilon$, we observe the perturbed solutions:
\begin{equation}
\label{Eq:subspace_IA_solution_perturb}
\varphi_1(t)\approx\varphi_3(t)=\psi_1,\quad 
\varphi_2(t)\approx-\varphi_4(t)=\psi_2. 
\end{equation}
Summarizing, the existence of IA patterns can be exactly proven for the limit $\varepsilon=0$. Since small $\varepsilon$ is a regular perturbation of system (\ref{Eq:general_model_4_IA_perturbation}), all asymptotically stable periodic attractors in this system will be only slightly perturbed by small $\varepsilon$-order terms, and one can observe patterns close to IA. 

\paragraph*{Numerical study of IA, AI, and AA patterns.}
The patterns IA, AI, and AA correspond to single lines of the order parameter distribution in Fig. \ref{Fig:EM}(C)-(E) and, hence, to  isolated attractors in the phase space. 
Figure~\ref{Fig:Attractor_IA_AI_AA} reports one example for each of these three patterns. 
All of them exhibit partial synchronization with quasiperiodic dynamics.

The phase-time plots in the column \textbf{b} of Fig.~\ref{Fig:Attractor_IA_AI_AA}(A) show that the 1st and 3rd pendulums are in-phase and the 2nd and 4th are anti-phase with $\varphi_1=\varphi_3$ and $\varphi_2=-\varphi_4$; this is also confirmed by the analytical solutions (\ref{Eq:subspace_IA_solution}) and (\ref{Eq:subspace_IA_solution_perturb}).
Inside the coupled groups, the 1st and 3rd or 2nd and 4th pendulums share the same mean frequency, leading to the multifrequency-clusters \cite{Berner2019b,omelchenko2011loss}.
Furthermore, the phase space projection on the plane $(\varphi_1,~\varphi_2)$ (orange curve) and the corresponding Poincaré map (black points, when $\varphi_4$ = 0 and $\dot{\varphi}_4 > 0$) indicates that the motion is quasiperiodic.
In summary, Fig.~\ref{Fig:Attractor_IA_AI_AA} reports different partially synchronous behaviors corresponding to isolated attractors with large-amplitude osillations: (A) IA, (B) AI, and (C) AA patterns. All these patterns coexist with EM contributing to a complex multistability scenario for the cross coupling structure (Fig.~\ref{Fig:Model}(G)).

\subsubsection{Three types of isolated attractors with small-amplitude oscillations}
The above-mentioned isolated attractors and EM states possess "large-amplitude" oscillations in the sense that their oscillation amplitude exceeds the threshold of the escapement mechanism $\varepsilon = 5.0^\circ$. 
This guarantees an inflow of energy into the system and the emergence of stable self-sustained oscillations of all clocks.

In this section, we investigate the case when some (or all) of the clocks in system (\ref{Eq:general_model_4}) do not reach this threshold. 
As a result, the SI, IS, and SS synchronization patterns appear (see Table \ref{Tab:phase possibility}). 
Figure~\ref{Fig:Attractor_Discontinuity} illustrates the numerically observed SI, IS, and SS patterns. For example, in Fig.~\ref{Fig:Attractor_Discontinuity}(B), both the time series and phase-time plots exhibit the in-phase synchronization between $\varphi_1$ and $\varphi_3$, while $\varphi_2$ and $\varphi_4$ converge to 0 and stop oscillating.
The periodic motions of $\varphi_1$ and $\varphi_3$ are illustrated by the phase trajectory (the orange line) and corresponding Poincaré map (black points, defined by $\varphi_1=0$ and $\dot{\varphi}_1 > 0$) in the column \textbf{d} of Fig.~\ref{Fig:Attractor_Discontinuity}(B).
Summarizing Fig.~\ref{Fig:Attractor_Discontinuity}, it shows the coexistence of three types of isolated attractors with small-amplitude oscillations, induced by the discontinuity of the escapement mechanism.

\begin{figure}[t] \centering
\includegraphics[width=0.31\textwidth]{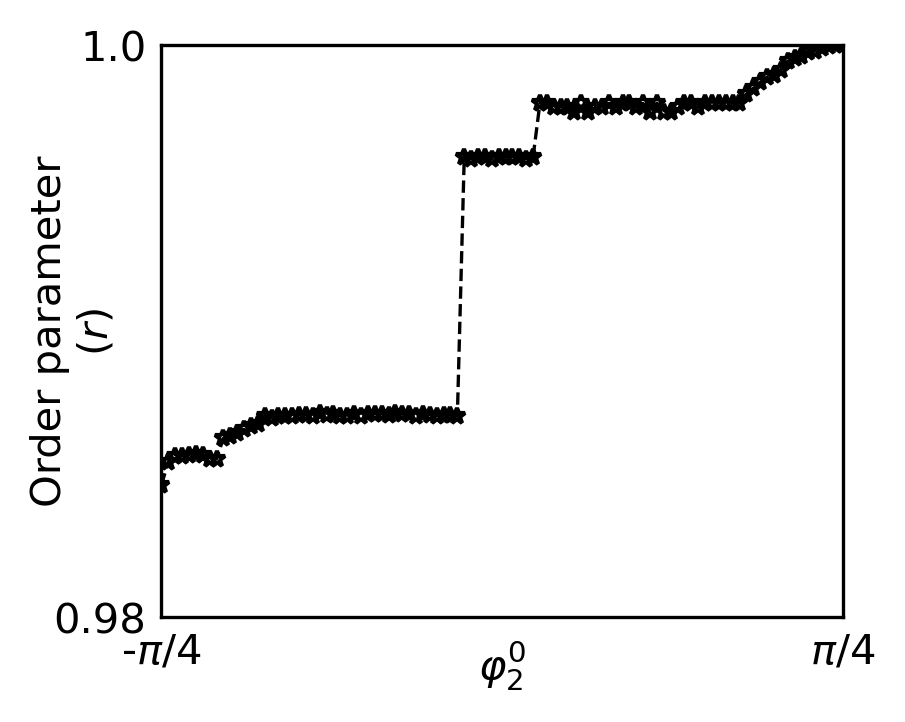}
\caption{
One-dimensional dependence of the order parameter on initial conditions for 4 coupled pendulums with cross-coupling structure (\ref{Eq:general_model_4}). 
The graph reveals the basins of different attractors along a one-dimensional line in the phase space. 
We fix 
[$\theta^0=0.01$, $\varphi_1^0=\frac{\pi}{4}$, $\varphi_3^0=\varphi_1^0+0.001$, $\varphi_4^0=\varphi_2^0+0.001$, $\dot\theta^0=\dot\varphi_1^0=\dot\varphi_2^0=\dot\varphi_3^0=\dot\varphi_4^0=0$],
and vary $\varphi_2^0$ to obtain 100 discretized values evenly distributed in the interval $[-\frac{\pi}{4},\frac{\pi}{4}]$.
The change of order parameter with respect to $\varphi_2^0$ indicates how synchronization patterns depend on initial conditions, including complete synchronization for $\varphi_2^0=\pi/4$.
Plateaus correspond to isolated attractors, continuously changing parts to EM states, and jumps to the basin boundaries.
Parameters are fixed as in Table \ref{Tab:Symbol} with $\alpha_1 = \frac{\pi}{2}$, $\alpha_2 = \pi$, $\alpha_3=\frac{3\pi}{2}$ and $\alpha_4=2\pi$ for the 4-coupled clocks ($N=4$).
}
\label{Fig:Basin_1D}
\end{figure}

\begin{figure}[!htbp] \centering
\includegraphics[width=0.26\textwidth]{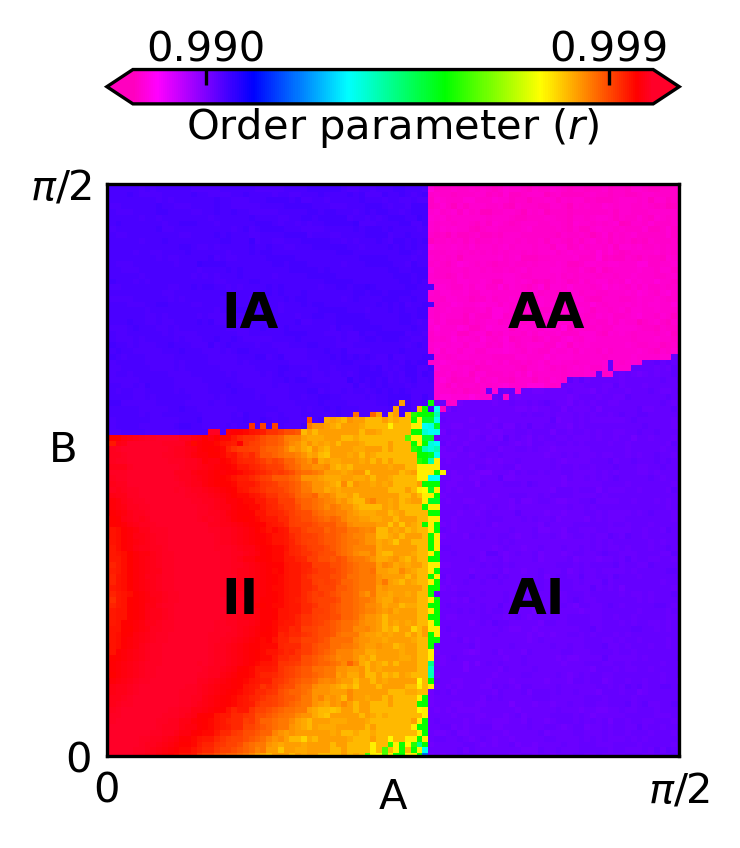}
\caption{
Two-dimensional basins of attraction of 4 coupled pendulums with cross-coupling structure (\ref{Eq:general_model_4}). The initial conditions are chosen as
[$\theta^0=0.01$, $\varphi_1^0=\varphi_2^0=\frac{\pi}{4}$, $\varphi_3^0=\varphi_1^0-A$, $\varphi_4^0=\varphi_2^0-B$, $\dot\theta^0=\dot\varphi_1^0=\dot\varphi_2^0=\dot\varphi_3^0=\dot\varphi_4^0=0$], with $A$ and $B$ 
changing in the interval $[0,\frac{\pi}{2}]$. The uniform discretization on the grid $100\times 100$ is used. 
Four regions in the bottom-left, top-left, bottom-right, and top-right correspond to the II,  IA, AI, and AA phase patterns, respectively.
The bottom-left part with the visible color gradient correspond to EM.
Parameters are fixed as in Table \ref{Tab:Symbol} with $\alpha_1 = \frac{\pi}{2}$, $\alpha_2 = \pi$, $\alpha_3=\frac{3\pi}{2}$ and $\alpha_4=2\pi$ for the 4-coupled clocks ($N=4$).
}
\label{Fig:Basin_2D}
\end{figure}

\begin{figure}[!htbp] \centering
\includegraphics[width=0.26\textwidth]{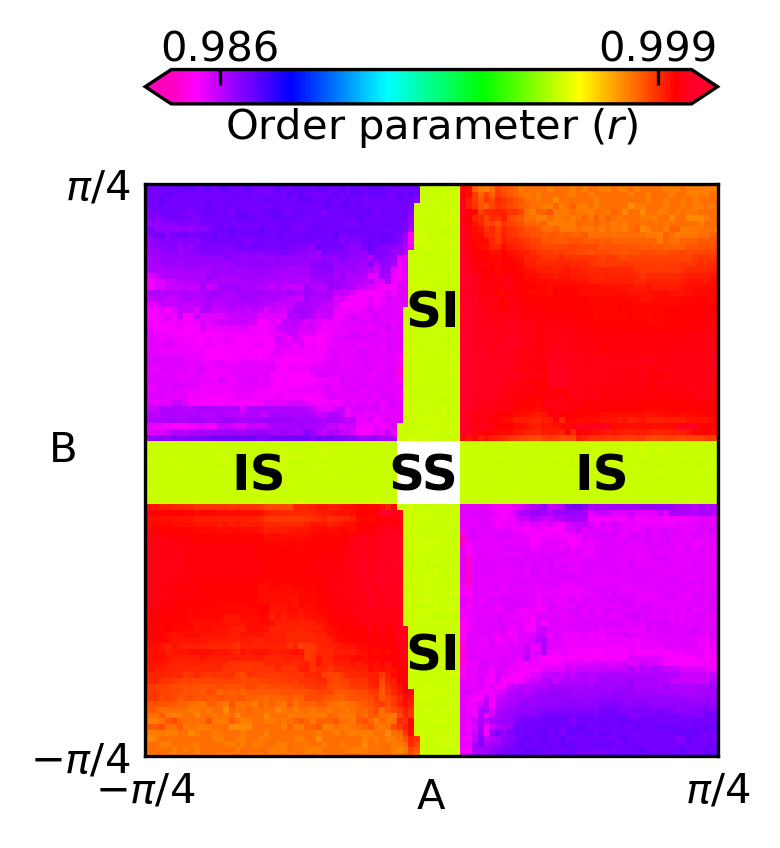}
\caption{
Two-dimensional basins of the attractors corresponding to the mixed-mode dynamics with one cluster staying silent and another oscillating. 
The model for 4 coupled pendulums with cross-coupling structure (\ref{Eq:general_model_4}) are considered. 
Initial conditions are chosen as follows: [$\theta^0=0.01$, $\varphi_1^0=\varphi_2^0=0$, $\varphi_3^0=A$, $\varphi_4^0=B$, $\dot\theta^0=\dot\varphi_1^0=\dot\varphi_2^0=\dot\varphi_3^0=\dot\varphi_4^0=0$], where $A$ and $B$ are taken from the interval $[-\frac{\pi}{4},\frac{\pi}{4}]$ discretized by $100\times 100$ evenly sampled points.
The white region corresponds to the trivial SS pattern (equilibrium at the origin); light green to SI and IS patterns.
Parameters are fixed as in Table \ref{Tab:Symbol} with $\alpha_1 = \frac{\pi}{2}$, $\alpha_2 = \pi$, $\alpha_3=\frac{3\pi}{2}$ and $\alpha_4=2\pi$ for the 4-coupled clocks ($N=4$).
}
\label{Fig:Basin_2D_Discontinuity}
\end{figure}

\section{Basins of attractions}\label{Sec:Basins of attractions}

Having clarified the different EM-related asymptotic phase patterns in 4 coupled clocks given by system (\ref{Eq:general_model_4}) in Sec.~\ref{Sec:Extreme multistability}, we discuss here their basins of attractions. 
Instead of randomly choosing initial conditions for each pendulum, we fix some of them while the rest are initialized with discretized values distributed evenly in the given intervals.
The corresponding synchronization states are estimated using order parameter $r$ from Eq. (\ref{Eq:order_parameter_used}), as it effectively identifies different attractors, including those belonging to EM.
Numerical results are summarized in Figs.~\ref{Fig:Basin_1D}, \ref{Fig:Basin_2D}, and \ref{Fig:Basin_2D_Discontinuity}. 

Figure~\ref{Fig:Basin_1D} shows the dependence of the order parameter $r$ 
on the initial conditions. 
For this we fix
[$\theta^0=0.01$, $\varphi_1^0=\frac{\pi}{4}$, $\varphi_3^0=\varphi_1^0+0.001$, $\varphi_4^0=\varphi_2^0+0.001$, $\dot\theta^0=\dot\varphi_1^0=\dot\varphi_2^0=\dot\varphi_3^0=\dot\varphi_4^0=0$],
and initialize $\varphi_2^0$ with 100 discretized values evenly distributed in the interval $[-\frac{\pi}{4},\frac{\pi}{4}]$.
We can clearly observe both the isolated attractors and a part of the EM regime.
Isolated attractors correspond to the flat segments.
For example, when $\varphi_2^0$ is around 0 in Fig. (\ref{Fig:Basin_1D}), in spite of different initial values of $\varphi_2^0$, trials in this segment part have almost the same $r$.
The continuously changing parts of $r$ in Fig. (\ref{Fig:Basin_1D}) are in line with the EM family of II states.
The abrupt "jumps" therefore represent boundaries between either the basins of the isolated attractors or between the isolated attractors and the EM family.

Figure \ref{Fig:Basin_2D} shows a two-dimensional basin of attraction, where we fix
[$\theta^0=0.01$, $\varphi_1^0=\varphi_2^0=\frac{\pi}{4}$, $\varphi_3^0=\varphi_1^0-A$, $\varphi_4^0=\varphi_2^0-B$, $\dot\theta^0=\dot\varphi_1^0=\dot\varphi_2^0=\dot\varphi_3^0=\dot\varphi_4^0=0$], and vary $A$ and $B$ in the interval $[0,\frac{\pi}{2}]$. 
The grid of $100 \times 100$ points is used.
One can observe four parts corresponding to II, IA, AI, and AA phase patterns.
In particular, the bottom-left part with the non-constant dependence of the order parameter $r$ in Fig.~\ref{Fig:Basin_2D} corresponds to a subset of an infinite number of stable II states from EM. 
The other three parts with constant colors are related to isolated attractors of AI, IA, and AA patterns.
With changing $A$ and $B$, the system moves from complete synchronization to inter-group anti-phase synchronization (see Fig.~\ref{Fig:Dynamical_Patterns_Of_EM}), and further to the multifrequency-cluster state (see Fig.~\ref{Fig:Attractor_IA_AI_AA}).

To visualize the basin of the attractors with small-amplitude oscillations, we analyse a set of initial conditions close to the origin. 
Figure~\ref{Fig:Basin_2D_Discontinuity} shows the corresponding basins of attraction, with the initial conditions
[$\theta^0=0.01$, $\varphi_1^0=\varphi_2^0=0$, $\varphi_3^0=A$, $\varphi_4^0=B$, $\dot\theta^0=\dot\varphi_1^0=\dot\varphi_2^0=\dot\varphi_3^0=\dot\varphi_4^0=0$], where $A$ and $B$ vary in the interval $[-\frac{\pi}{4},\frac{\pi}{4}]$. In addition to II patterns from EM set, we obtain here three new regions corresponding to the patterns SI, IS, and SS, for which some of the clocks are not oscillating. 
The basin of the trivial solution SS with all clocks silent is observed in the central part (white); see Fig.~\ref{Fig:Attractor_Discontinuity}(C) for the illustration of the pattern.
The SI and IS patterns are mixed-mode oscillations, with one cluster silent and another cluster oscillating, corresponding to the light green basin of attraction.

\section{Conclusions}\label{Sec:Conclusion}
In summary, we investigate how different coupling topology affects the collective dynamics in coupled clocks. 
The considered model includes global as well as local couplings represented by the rotating support disc and springs, respectively. 
Also, the model contains a discontinuity due to the escapement clock mechanism.
We focus on the model of 4 coupled clocks where surprisingly an EM is observed between patterns with different synchronization levels. 
The EM phenomenon reveals the coexistence of infinitely many stable asymptotic states.
The dependence on initial conditions is clarified using the analysis of the basins of attractions.
The main conclusions based on our work are as follows:
\begin{itemize}
\item For both three coupled and four coupled clocks, we use Monte Carlo sampling and find that the symmetric coupling structure can increase the dynamical complexity.
This can lead to diverse synchronization patterns and attractors.
\item We observe the emergence of EM for the case of four coupled clocks (see Fig. \ref{Fig:Model}(G) and model (\ref{Eq:general_model_4})).
This phenomenon is solely induced by the cross-coupling topological structure, and it is stable against variations of the system parameters, as far as the clocks remain identical.
\item We show both analytically and numerically that the emergence of EM is closely related to the II synchronization pattern (see Table \ref{Tab:phase possibility}), where the system splits into two antipodal clusters such that the clocks within the clusters are fully synchronized, but the intra-cluster dynamics can be shifted by an arbitrary phase. 
Moreover, other three types of isolated attractors with large-amplitude oscillations coexist with the infinite family of states from the EM family.
They correspond to the IA, AI, and AA synchronization patterns.
\item We further uncover the effect of the discontinuity of the system induced by the escapement mechanism. 
It induces the emergence and coexistence of further three types of isolated attractors with small-amplitude oscillations. These states  correspond to SI, IS, and SS patterns (see Table \ref{Tab:phase possibility}).
In particular, the IS and SI patterns are mixed states \cite{Ebrahimzadeh2022} where two clocks are oscillating and the other two stay silent.  
\end{itemize}

As observed in our work, the emergence of EM is caused by a particularly designed symmetric coupling structure, rather than by introducing additional quantities into the coupling design.
The inclusion of more coupled clocks with different topological coupling structures is an open way to clarify the emergence of EM or chimera states of large coupled systems.
Another possible generalization is to include adaptivity in the coupling scheme.

\section*{Acknowledgments}
Z.S. and Y.-R.L. was funded by the China Scholarship Council (CSC) scholarship.
J.K. was supported by the Federal Ministry of Education and Research (BMBF) grant No. 01LP1902J (climXtreme).
S.Y. was supported by the German Research Foundation DFG, Project No. 411803875.

\section*{Author Contributions}
Z.S. and Y.S. contributed equally to this work.

\section*{Data Availability}
The code used for this work is available online \href{https://github.com/Zsstarry/EM\_Clocks}{here}.

\bibliography{EM_Clocks}
\end{document}